\documentclass[12pt,preprint]{aastex}

\usepackage{epsfig}
\usepackage{amsmath}

\begin{document}

\title{{\em Spitzer} Observations of the Brightest Galaxies \\
in X-Ray-Luminous Clusters}

\author{
E.~Egami\altaffilmark{1},
K.~A.~Misselt\altaffilmark{1},
G.~H.~Rieke\altaffilmark{1},
M.~W.~Wise\altaffilmark{2},
G.~Neugebauer\altaffilmark{1},
J.-P.~Kneib\altaffilmark{3,4},
E.~Le~Floc'h\altaffilmark{1,5},
G.~P.~Smith\altaffilmark{4,6},
M.~Blaylock\altaffilmark{1},
H.~Dole\altaffilmark{1,7},
D.~T.~Frayer\altaffilmark{8},
J.-S.~Huang\altaffilmark{9},
O.~Krause\altaffilmark{1},
C.~Papovich\altaffilmark{1},
P.~G.~P\'{e}rez-Gonz\'{a}lez\altaffilmark{1},
and
J.~R.~Rigby\altaffilmark{1}
}

\altaffiltext{1}{Steward Observatory, University of Arizona, 933
  N. Cherry Avenue, Tucson, AZ~85721}
\altaffiltext{2}{Massachusetts Institute of Technology, Center for
Space Research, Building 37, 70 Vassar Street, Cambridge, MA~02139-4307}
\altaffiltext{3}{OAMP, Laboratoire d'Astrophysique de Marseille,
UMR6110 traverse du Siphon, 13012 Marseille, France}
\altaffiltext{4}{Division of Physics, Mathematics, and Astronomy,
California Institute of Technology, 105-24, Pasadena, CA~91125}
\altaffiltext{5}{Associated to Observatoire de Paris, 92195 Meudon, France}
\altaffiltext{6}{School of Physics and Astronomy, University of
Birmingham, Edgbaston, Birmingham, B15 2TT, England}
\altaffiltext{7}{Institut d'Astrophysique Spatiale, b\^{a}t 121,
  Universit\'{e} Paris Sud, F-91405 Orsay Cedex, France}
\altaffiltext{8}{Spitzer Science Center, California Institute of
  Technology 220-06, Pasadena, CA~91125}
\altaffiltext{9}{Harvard-Smithsonian Center for Astrophysics, 60
  Garden St., Mailstop 65, Cambridge, MA~02138}

\begin{abstract}

We have studied the infrared properties of the brightest cluster
galaxies (BCGs) located in the cores of X-ray-luminous clusters at
$0.15 < z < 0.35$.  The majority of the BCGs are not particularly
infrared-luminous compared with other massive early-type galaxies,
suggesting that the cluster environment has little influence on the
infrared luminosities of the BCGs.  The exceptions, however, are the
BCGs in the three X-ray-brightest clusters in the sample, A1835,
Z3146, and A2390.  These BCGs have a prominent far-infrared peak in
their spectral energy distributions (SEDs), and two of them (those in
A1835 and Z3146) can be classified as luminous infrared galaxies
(LIRGs: $ L_{IR} >10^{11}$ L$_{\sun}$).  Although radio AGNs are found
to be prevalent among the BCGs, the infrared luminosities of these
three BCGs, judged from the infrared SED signatures, are likely to be
powered by star formation.  Considering the overall trend that clusters
with shorter radiative gas cooling times harbor more infrared-luminous
BCGs, the enhanced star formation may be caused by the cooling cluster
gas accreting onto the BCGs.

\end{abstract}


\section{Introduction}

Clusters of galaxies are embedded in a halo of diffuse hot gas
($T_{gas} \sim 10^{7}-10^{8}$ K), which emits strongly in the X-ray
through thermal bremsstrahlung radiation.  The radiative cooling time
of this hot intracluster plasma is usually comparable to the Hubble
time, suggesting that the gas does not cool substantially over the age
of the clusters.  There are, however, clusters that show a highly
peaked X-ray surface brightness profile in the core, which indicates
that the hot gas in these cores is dense and has a substantially
shorter cooling time.  In extreme cases, the cooling time is estimated
to be less than a few Gyrs, which prompted the idea that these dense
cores are sustaining an inflow of cooling gas toward the cluster
center \citep{Cowie77,Fabian77}.

The reality of such a cluster ``cooling flow'' has been disputed over
the years on the basis that the inferred gas cooling rate (which we
assume to lead to a mass deposition rate) is simply too large to be
consistent with the modest level of activity (e.g., H$\alpha$-derived
star formation rates) seen in these cluster cores (see
\citet{Fabian94} for a review of cooling flows).  In the most extreme
cases, the gas cooling rate was estimated to be $>$1000 M$_{\sun}$
yr$^{-1}$ \citep[e.g.,][]{Allen96}, and yet there is no sign of
activity involving such a magnitude of mass inflow.  Many of the
brightest cluster galaxies (BCGs) in strongly cooling cores do show
some signs of activity, such as visual emission lines and UV continuum
\citep[e.g.,][]{Allen95} , but if they are due to star formation, the
inferred star formation rate (e.g., from H$\alpha$) is only a few
hundred M$_{\sun}$ yr$^{-1}$ at most \citep[e.g.,][]{Crawford99}.
Considering that the BCGs in strongly cooling cores also show evidence
for containing a significant amount of dust \citep{Edge99} and
molecular gas \citep{Edge01,Salome03}, it is clear that these BCGs are
unusually active; what is puzzling, however, is the large difference
between the gas cooling rate we infer and that we can account for.

A self-consistent picture is finally emerging from observations with
{\em Chandra} and {\em XMM}.  {\em XMM} especially has demonstrated
convincingly for the first time that the cluster gas is cooling but
not as much as originally thought.  More specifically, {\em XMM} has
detected strong emission from a cool plasma at temperatures half to
one-third of the ambient value, but there is also a severe deficit of
emission from plasma below these temperatures
\citep{Peterson01,Tamura01,Kaastra01,Peterson03}.  In other words,
there seems to be a floor of gas temperature below which cooling is no
longer effective.  The result is an order of magnitude reduction in
the X-ray derived mass deposition rates, making them consistent with
the modest amount of activity observed at other wavelengths, such as
the star formation rates derived from the extinction-corrected
H$\alpha$ emission \citep{Crawford99}.

The current view is that some kind of gas heating (e.g., a radio jet
from an AGN) slows the cooling process (see \citet{Begelman04} for a
recent review).  Indeed, such heating may play an important role in
the process of galaxy formation in general because it would provide a
way to suppress the formation of bright ($ > L^{*}$) massive galaxies,
which are overproduced in model calculations (\citet{Croton06} and
references therein).  In fact, the BCGs in the cooling-flow clusters
may prove to be an important laboratory where we can study the effects
of cooling gas directly accreting onto a seed mass concentration, a
condition that might be found in the first generation of forming
galaxies at high redshift.

To investigate this cooling flow problem further, we have examined the
{\em Spitzer}/IRAC (3.6, 4.5, 5.8, and 8.0 $\mu$m) and {\em
Spitzer}/MIPS (24, 70, and 160 $\mu$m) images of the BCGs located in
the cores of X-ray-luminous clusters.  At the redshifts of targeted
clusters, IRAC detects stellar continuum light while
MIPS detects thermal dust emission.  Observations of emission from
heated dust can detect deeply embedded forms of activity that are
hidden from other search methods such as visual emission lines or UV
continuum.  For example, a galaxy with a star formation rate of $\sim
1000$ M$_{\sun}$ yr$^{-1}$ would be ultraluminous in the infrared
(L$_{\rm IR} > 10^{12}$ L$_{\sun}$), so if any of the BCGs in the
so-called cooling flow clusters are undergoing such vigorous star
formation as a result of a massive gas inflow, they would stand out in
the {\em Spitzer} mid-/far-infrared images.  The overall shape of
spectral energy distribution (SED) will also tell us whether the
infrared luminosity source is star formation or an AGN.

Throughout the paper, we adopt the cosmological parameters of
$\Omega_{M}=0.3$, $\Omega_{\Lambda}=0.7$, and $H_{0}=70$ km s$^{-1}$
Mpc$^{-1}$ except when noted otherwise. 

\section{The Sample}

The eleven clusters for this study are listed in Table~\ref{flux} and
were part of a {\em Spitzer} GTO program described in \citet{Egami05}.
The overall cluster selection criteria for this program are the
following: (1) X-ray luminous ($L_{X} \ga 5 \times 10^{44}$ erg/s
based on the published soft X-ray luminosities listed in
Table~\ref{flux}), (2) moderate redshift ($z \sim 0.15-0.5$), and (3)
low IR background ($N_{H} < 3.5 \times 10^{22}$ cm$^{-2}$).  We did
not include clusters whose redshifts are based on a small number of
measurements and those with severely limited ancillary data.  Since
the primary goal of this program is to study gravitationally lensed
background galaxies, the most X-ray-luminous clusters were chosen
under the assumption that they are the most massive clusters and
therefore the most effective lenses.  This means that these clusters
were chosen without any regard to the properties of the BCGs.
However, since, to a first order, the X-ray luminosity is proportional
to the gas cooling rate, this cluster sample also provides a good data
set to study the properties of BCGs in strongly cooling cluster cores.

The identification of the BCG in a cluster is usually straightforward.
Many cluster cores harbor a single giant elliptical galaxy that
dominates the brightness at the visual/near-infrared wavelengths and is
located at/near the peak of the cluster X-ray emission.  There are,
however, cases in which a BCG cannot be identified uniquely: some
clusters have more than one giant elliptical with similar brightnesses
in the cluster core, or have a disturbed X-ray morphology without a
well-defined peak.  In these cases, it was not clear how we could
relate the cluster gas properties (e.g., cooling rates) with those of
the individual galaxies.  For this study, we therefore chose only
those clusters that show an X-ray morphology with a well-defined
single peak and have a single 24 $\mu$m-detected BCG near the X-ray
peak based on the archived ROSAT HRI images and on published {\em
Chandra} images \citep[e.g.,][]{Smith05}.  We also note that based on
the {\em Chandra} images, our sample contains both ``relaxed'' and
``unrelaxed'' clusters, and that the spatial coincidence of the X-ray
peak and BCG is not precise in many of the ``unrelaxed'' clusters
\citep[e.g.,][]{Smith05}.

Table~\ref{flux} lists the position of the BCG in each cluster as well
as a few critical properties of these clusters such as X-ray
luminosities and gas cooling times.  Based on the classification by
\citet{Allen98}, 7 out of the 11 clusters are classified as cooling
flow clusters (CF) while four are classified as non-cooling flow
clusters (NCF).

\section{The Data}

\subsection{{\em Spitzer}/IRAC Data}

The 3.6, 4.5, 5.8, and 8.0 $\mu$m images were obtained using the
Infrared Array Camera (IRAC; Fazio et al. 2004) on board the {\em
Spitzer} Space Telescope \citep{Werner04}.  Each of the four IRAC
channels uses a separate detector array.  The 3.6 $\mu$m channel
($\lambda_{c}=3.56 \mu$m; $\Delta\lambda = 0.75 \mu$m) and 4.5 $\mu$m
channel ($\lambda_{c}=4.52 \mu$m; $\Delta\lambda = 1.01 \mu$m) use
$256 \times 256$ InSb arrays while the 5.8 $\mu$m channel
($\lambda_{c}=5.73 \mu$m; $\Delta\lambda = 1.42 \mu$m) and 8.0 $\mu$m
channel ($\lambda_{c}=7.91 \mu$m; $\Delta\lambda = 2.93 \mu$m) use
$256 \times 256$ Si:As arrays.  At any given time, two channels are
used simultaneously (3.6 and 5.8 $\mu$m; 4.5 and 8.0 $\mu$m).  Each
channel has a similar pixel size ($\sim$1\farcs2 pixel$^{-1}$) and
field of view (5\farcm2$\times$5\farcm2).  Integration times were
2400~s per band except for A1835, whose integration times were 2400~s
at 3.6 \& 5.8 $\mu$m and 3600~s at 4.5 \& 8.0 $\mu$m\footnote{The IRAC
data of A1835 were taken as part of the GTO program PID:64 (PI:
Giovanni Fazio).  All the other data were taken as part of the GTO
program PID:83 (PI: George Rieke).}.  The images were taken with the
small-step cycling dither pattern.  The basic calibrated data (BCD)
images were combined using a custom IDL mosaicking routine with the
final pixel scale of 0\farcs6 pixel$^{-1}$, half of the instrument
intrinsic pixel size.

The IRAC photometry used a circular beam with a diameter of 12\farcs2
with a sky background annulus of 6\farcs1--12\farcs2 in radius.  We
selected this rather large beam (6$\times$ the PSF FWHM) because at
these wavelengths, some BCGs show a significant spatial extent.
Although in some cases there is contamination from nearby sources
falling into the beam, the effect is small because of the dominance of
the BCGs.  The point-source aperture corrections were applied, which
were 1.061, 1.064, 1.067, and 1.089 at 3.6, 4.5, 5.8, and 8.0 $\mu$m,
respectively, based on the IRAC Data Handbook.  

The resultant 1 $\sigma$ sensitivities in the mosaicked images are
$\sim 5 \mu$Jy in all four bands.  The shorter-wavelength images do not
necessarily go deeper when such a large photometry beam is used with
IRAC images of this depth due to the confusion noise.  A conservative
estimate for the absolute calibration uncertainty is 10\%.

\subsection{{\em Spitzer}/MIPS Data}

The 24, 70, and 160 $\mu$m images were obtained using the Multi-band
Imaging Photometer for {\em Spitzer} (MIPS; Rieke et al. 2004).  The
MIPS 24$\mu$m channel ($\lambda_{c}=23.7 \mu$m; $\Delta\lambda = 4.7
\mu$m) uses a 128$\times$128 BIB Si:As array with a pixel scale of
2\farcs55 pixel$^{-1}$.  The 70 $\mu$m channel ($\lambda_{c}=71.4
\mu$m; $\Delta\lambda = 19.0 \mu$m) uses a 16$\times$32 Ge:Ga array
with a pixel scale of 9\farcs98 pixel$^{-1}$.  The 160 $\mu$m channel
($\lambda_{c}=155.9 \mu$m; $\Delta\lambda = 35 \mu$m) uses a
2$\times$20 stressed Ge:Ga array with a pixel scale of
16\arcsec$\times$18\arcsec.

All the clusters in the sample (Table~\ref{flux}) were imaged at 24
$\mu$m using the large-source photometry mode.  The total integration
times were 3600 s pixel$^{-1}$ at the position of the BCGs.  Since we
did not have enough observing time to observe all the BCGs at 70 and
160 $\mu$m, at these wavelenghts we only observed the three BCGs
brightest at 24 $\mu$m, for which we can expect a good chance of
detection.  For the two clusters with the brightest BCGs at 24 $\mu$m
(A1835 and Z3146), we conducted 70 and 160 $\mu$m imaging using the
large-source photometry mode.  For the cluster with the third
brightest BCG at 24~$\mu$m (A2390), we performed only 70 $\mu$m
imaging due to a high infrared sky background toward this direction.
The integration times were 370 s pixel$^{-1}$ at 70 $\mu$m, and 100 s
pixel$^{-1}$ at 160 $\mu$m at the position of the BCGs.

The data were reduced and combined with the Data Analysis Tool (DAT)
developed by the MIPS instrument team \citep{Gordon05}.  A few
additional processing steps were applied to the 24 $\mu$m data as
well.  First, special flat fields were constructed to remove artifacts
in the flats produced by material deposited on the scan
mirror. Secondly, the 24~$\mu$m array is subject to latent images
after exposure to a bright source. The latent images manifest
themselves as a region of reduced sensitivity centered on the array
position of the bright source in subsequent exposures.  These ``dark
latents'' decay with time and disappear upon the next anneal cycle.
Several of our observations were contaminated by dark latents caused
by bright 24 $\mu$m sources in the target fields.  A source-free image
of the latents was produced by median combining the entire data set
with appropriate bright source masks.  The dark latents were removed
from each individual frame by dividing the frames by the normalized
dark latent image prior to mosaicking. A detailed discussion of these
instrumental effects will be presented in a forthcoming paper
(G.~H.~Rieke et al.\ 2006, in preparation).  A third processing step
beyond the normal DAT processing was necessary owing to the design of
the observations.  Each cluster was observed using two Astronomical
Observing Requests (AORs), and the two AORs were often separated by a
significant amount of time.  As a results, changing zodiacal
backgrounds resulted in strong gradients in the resulting mosaics.  To
mitigate the effect of any DC offset between AORs, a mean row was
subtracted from each individual data frame, again with an appropriate
bright source mask, prior to mosaicking.  The final mosaicked images
were produced with pixel scales of 1\farcs25, 4\farcs92, and 8\arcsec\
pixel$^{-1}$ at 24, 70, and 160 $\mu$m, respectively, which are half
of the instrument intrinsic pixel sizes.

The MIPS photometry beam diameters and corresponding inner/outer sky
annulus radii are as follows: 26\arcsec\ and 20\arcsec--32\arcsec\ at
24 $\mu$m, 35\arcsec\ and 39\arcsec--65\arcsec\ at 70 $\mu$m, and
50\arcsec\ and 75\arcsec--125\arcsec\ at 160 $\mu$m.  The point-source
aperture corrections for these measurements are 1.167, 1.308, and
1.442, respectively, based on the MIPS instrument web page.  For the
clusters A2218 and AC114, we used a smaller photometry beam (13\farcs2
diameter, aperture correction $=$ 1.648) at 24 $\mu$m to prevent
contamination from nearby sources.  The 70 and 160 $\mu$m aperture
corrections depend to some extent on the intrinsic color of the
observed source, so we used the value for a red PSF ($T=15$ K) at 70
$\mu$m and that for a blue PSF ($T=1000$ K) at 160 $\mu$m, assuming
that the infrared SEDs of the galaxies observed here are peaking
somewhere between these two wavelengths.  However, the effect of this
color-dependent aperture correction is on the order of a few to
several percent at most, well within the uncertainties of our
photometric measurements.

The resultant 1 $\sigma$ sensitivities in the mosaicked images are
0.01, 2.6, and 15 mJy at 24, 70, and 160 $\mu$m, respectively.  The
absolute calibration uncertainties are 10, 20, and 20\% in these three
bands.

\subsection{{\em Chandra} X-ray Data}

For the three X-ray brightest clusters in the sample (A1835, A2390,
and Z3146), we have extracted all available data from the {\em Chandra}
archive, and have reprocessed them to provide a uniform and consistent
set of X-ray measurements.  The initial {\em Chandra} results for A1835 have
been published previously in \citet{Schmidt01}.  For A2390, an
analysis of the combined 19~ksec observation from Cycle 1 has been
published in \citet{Allen01}; however, a subsequent deeper exposure
was taken in Cycle 4, and we also included this data set.  Finally,
although observed during Cycle 1, the {\em Chandra} results for Z3146 remain
unpublished.  When all the data were combined, the total integration
times for A1835, A2390, and Z3146 (after standard filtering) were 30
ksec (with two data sets), 104 ksec (with three data sets), and 46
ksec (with one data set), respectively.  The data processing and the
method for calculating the gas cooling rates are described in
Appendix~\ref{chandra}.  Detailed analyses of the A2390 and ZW3146
{\em Chandra} data will be published elsewhere (M.~W.~Wise 2006, in
preparation).

\subsection{{\em VLA} 1.4 GHz Data}

We have compiled the 1.4 GHz flux densities of the BCGs from a number
of sources.  Where duplications exist, we have adopted the following
priorities:
\begin{enumerate}
  \item Deep high-resolution observations targeting clusters
    \citep{Morrison03, Rizza03}
  \item FIRST survey (2003 Apr 11 version) \citep{Becker95,White97}
  \item NVSS survey \citep{Condon98}
\end{enumerate}
The data published by \citet{Morrison03} and \citet{Rizza03} have the
highest spatial resolution (1\farcs5, A configuration) and sensitivity
($\sigma \sim 25-50 \mu$Jy).  The FIRST and NVSS surveys have spatial
resolutions of 5\arcsec\ (B configuration) and 45\arcsec\ (D
configuration), and sensitivities of $\sigma \sim 140$ and 450
$\mu$Jy, respectively.  The flux densities from the NVSS should be used
with caution because we see cases in which a single NVSS source at
a BCG position breaks up into multiple radio components (e.g., Z3146)
or is clearly displaced from the BCG (e.g., A2219) when observed with higher
spatial resolution.

All the VLA flux densities are listed in Table~\ref{spitzer}.  When
there is no radio detection (i.e., A773, A2218, and A2219), we list a
3 $\sigma$ upper limit.

\section{Results}

\begin{figure}

  \hspace*{-1cm}\includegraphics[angle=90,scale=0.8]{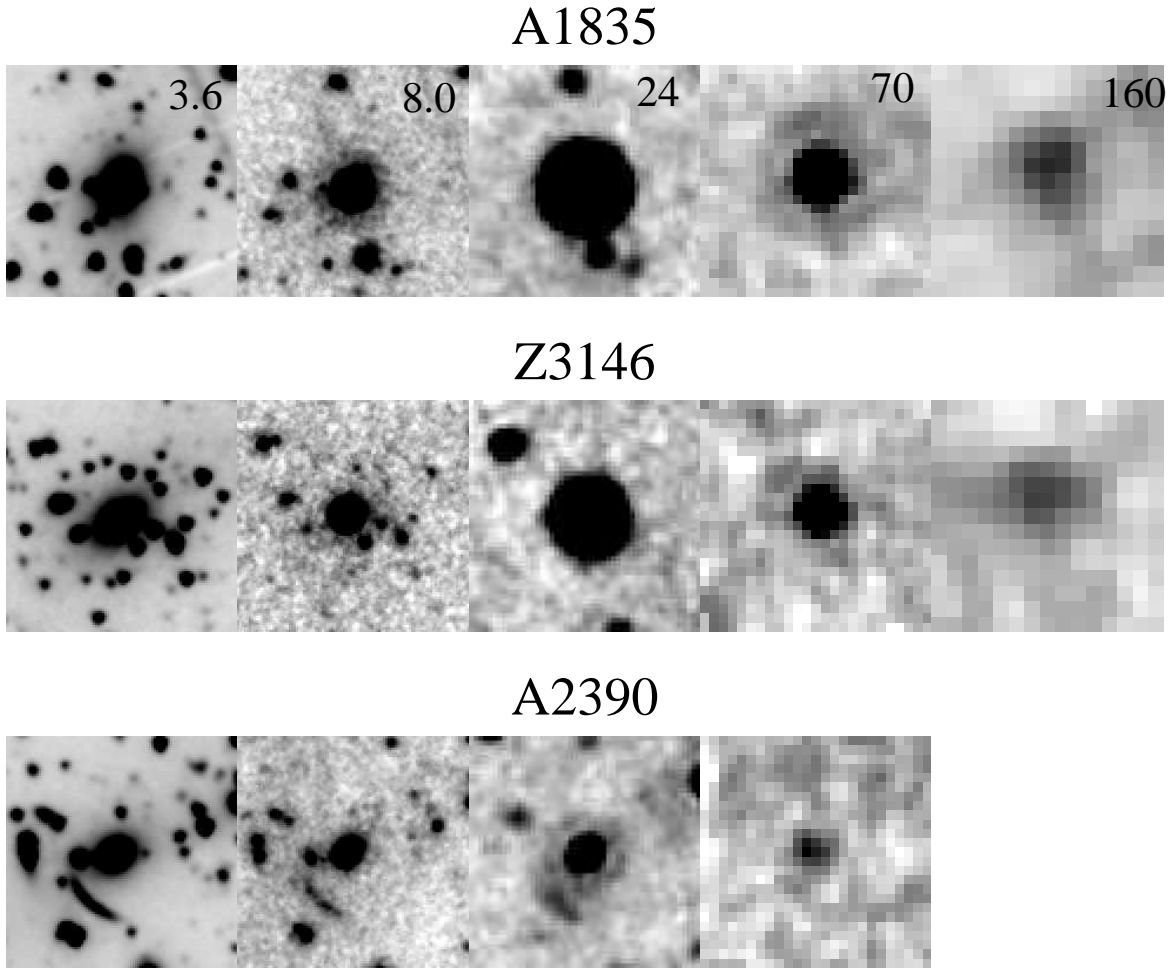}
  
  \caption{{\em Spitzer} images of the cluster cores harboring the
  three infrared-brightest BCGs.  From left to right are the 3.6, 8.0,
  24, 70, and 160 $\mu$m images, and the clusters are ordered
  vertically in decreasing 24 $\mu$m flux densities.  North is up and
  east is left.  Each image is 1\arcmin\ on a side, and the pixel
  scales are 0\farcs6 pixel$^{-1}$ at 3.6 and 8.0 $\mu$m, 1\farcs25
  pixel$^{-1}$ at 24 $\mu$m, 4\farcs92 pixel$^{-1}$ at 70 $\mu$m, and
  8\arcsec\ pixel$^{-1}$ at 160 $\mu$m.
  \label{three_brightest}}
\end{figure}

\begin{figure}
  \hspace*{1cm}\includegraphics[angle=90,scale=0.6]{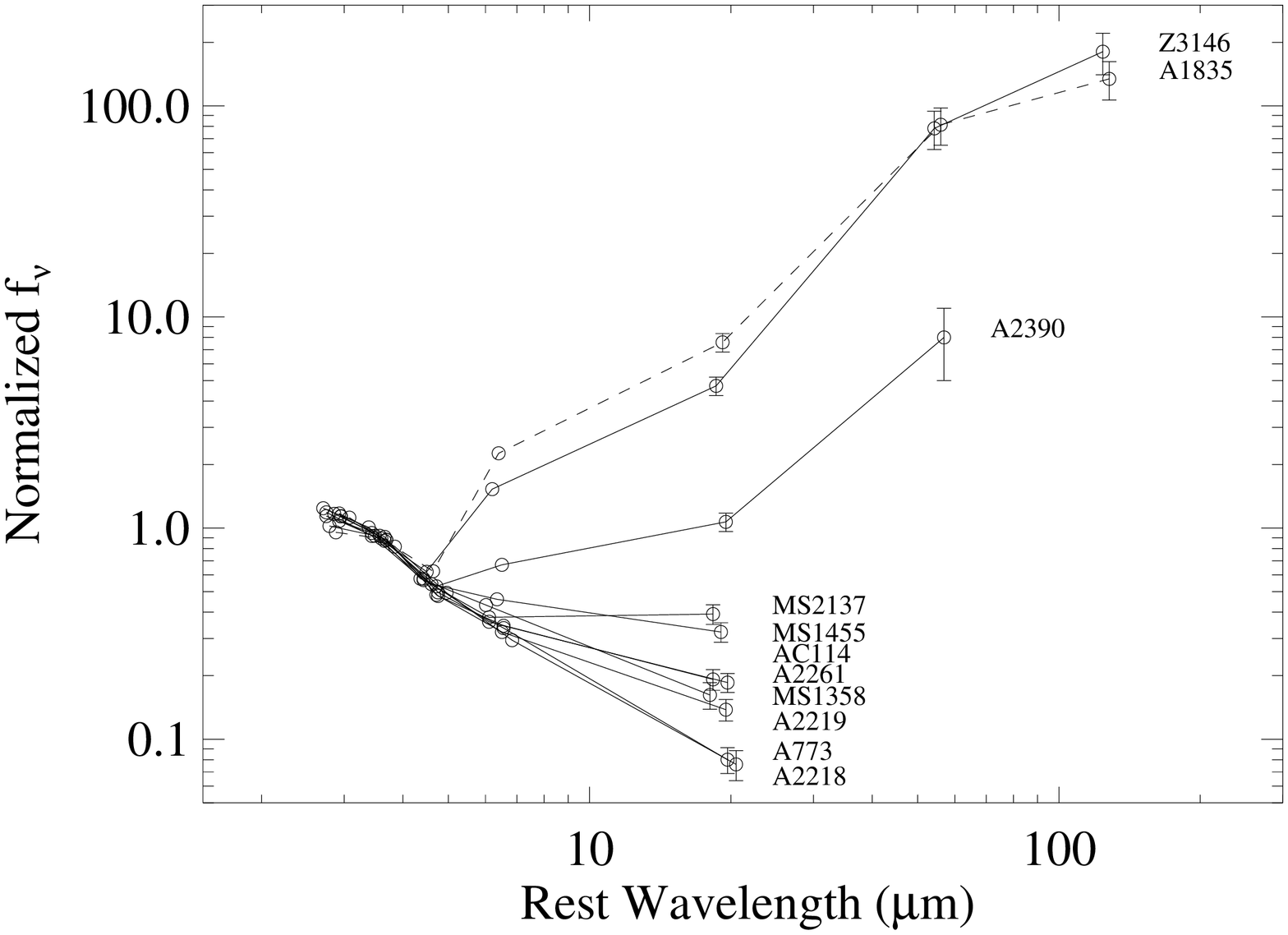}

  \caption{{\em Spitzer} SEDs of the BCGs.  All the SEDs were scaled
  such that the flux densities coincide around 4.5 $\mu$m. The SED of
  the BCG in A1835 is denoted as a dashed line to make it
  distinguishable from that of Z3146. \label{sed}}
\end{figure}

Figure~\ref{three_brightest} and \ref{24um_image} show the IRAC and
MIPS images of the core regions of the eleven clusters.  The measured
{\em Spitzer} flux densities are given in Table~\ref{spitzer}.
Figure~\ref{three_brightest} shows the {\em Spitzer} images of the
three BCGs that are the brightest at 24 $\mu$m.
Figure~\ref{24um_image} shows the rest of the sample.  Note that these
three BCGs are also the brightest in terms of the intrinsic
mid-infrared luminosity since all the clusters studied here are at
similar redshifts ($z \sim 0.2-0.3$).  The SEDs produced from the
IRAC/MIPS photometry are shown in Figure~\ref{sed}\footnote{In this
and subsequent figures in the paper, error bars are omitted when their
sizes are comparable to or smaller than those of the plot symbols.}.

The strong infrared sources in these three BCGs are compact, the 24
$\mu$m sources essentially being unresolved, which sets an upper limit
on the source diameter ($D$) of $D < 20$ kpc for the size of the
infrared emitting region.  The BCGs are resolved at 8 $\mu$m with an
estimated intrinsic size of $D \sim 8$ kpc, but the stellar light
contribution (as opposed to thermal dust emission) may be significant
at this wavelength.  Therefore, we treat this size as an upper limit.
This indicates that the infrared luminosity originates in the BCGs.

The rest of the BCGs in the sample show SEDs monotonically decreasing
in the mid-infrared with no clear indication of having a significant
amount of far-infrared luminosity.  These SEDs could still peak in the
far-infrared, but the corresponding far-infrared luminosities would be
much smaller than those of the three infrared-bright BCGs.  At the
same time, all the BCGs are detected above 5 $\sigma$ at 24 $\mu$m,
indicating that these BCGs are not completely infrared-quiet.

Figure~\ref{sed} also shows that none of the BCGs has a power-law SED
in the infrared typical of an AGN-dominated galaxy.  Instead, all of
them show a gradual drop at $\lambda_{\rm rest} \sim 3-5 \mu$m, a
characteristic of star-dominated SED (here, we see only the longer
wavelength drop from the SED peak near 1.6 $\mu$m in the restframe).
In fact, despite the large variation in the mid-/far-infrared range,
the SEDs of the BCGs are quite similar at $\lambda_{\rm rest} \sim 3-5
\mu$m not only in terms of shape but also of luminosity.  This
indicates that the near-infrared light of these BCGs is dominated by
equally massive stellar populations, which are not directly related to
the generation of the infrared luminosity.

\section{Discussion}

\subsection{Infrared Properties of the BCGs}

Figure~\ref{plot_24_x} plots the restframe 15 $\mu$m monochromatic
luminosities of the BCGs against the soft X-ray (0.1--2.4 keV)
luminosities of their parent clusters.  The soft X-ray luminosity is a
good measure of the intracluster gas cooling rate, and unlike the hard
X-ray, it will not be affected severely by AGN emission.
Table~\ref{flux} lists the {\em ROSAT} soft X-ray (0.1--2.4 keV)
luminosities from two {\em ROSAT} cluster catalogs, the {\em ROSAT}
Brightest Cluster Sample (BCS) \citep{Ebeling98} and the Northern {\em
ROSAT} All-Sky (NORAS) Galaxy Cluster Survey \citep{Boehringer00}.  As
seen in Table~\ref{flux}, the luminosities listed in these two
catalogs are consistent with each other with a $\sim$10\% uncertainty;
we used the latter catalog by default.  The two southern clusters,
MS2137.3$-$2353 and AC114, are not listed in the {\em ROSAT}-ESO Flux
Limited X-ray (REFLEX) Galaxy Cluster Survey \citep{Boehringer04}, the
southern-sky version of NORAS, so instead we used the Einstein
0.3--3.5 keV \citep{Gioia94} and ASCA 2--10 keV \citep{Allen00} X-ray
luminosities, respectively, as surrogates for the ROSAT soft X-ray
luminosities.  The latter needs to be divided by a factor of 1.25 to
account for the systematic offset between the ASCA and ROSAT
luminosities\footnote{Although the {\em ASCA} passband covers the hard
X-ray and not soft X-ray, Table~\ref{flux} shows that emperically
there is a good scaling relation between the {\em ASCA} and {\em
ROSAT} luminosities in the sense that the former is $1.25\pm0.05$
times larger than the latter when A2219 and A2390 are exluded.  The
hard X-ray luminosities of these two clusters are likely boosted by
AGN emission.  As we will show later, the BCG in A2390 contains a
strong radio AGN while A2219 is known to have a strong radio source in
the cluster core near the BCG.}.

We have estimated the restframe 15 $\mu$m monochromatic luminosities
based on the SEDs shown in Figure~\ref{sed}.  We chose to calculate
the monochromatic luminosity at 15 $\mu$m because the conversion from
the 15 $\mu$m luminosity to the total infrared luminosity is well
established based on the {\em ISO} observations \citep{Elbaz02}.  For
comparison, the 1 $\sigma$ range of the 15 $\mu$m luminosities
measured for local early-type galaxies by ISO \citep{Ferrari02} is
shown (the two horizontal dotted lines).  The figure shows that the
majority of the BCGs studied here are not particularly luminous in the
mid-infrared compared with the sample of \citet{Ferrari02}.  Ferrari
et al.'s sample may be biased toward infrared-active early-type
galaxies, but this comparison at least shows that there is nothing
special about the infrared luminosities of most BCGs, suggesting that
the cluster environment has little influence on the BCG infrared
properties in general.

The only exceptions are the BCGs in A1835, Z3146, and possibly A2390.
The restframe 15 $\mu$m luminosities of the BCGs in A1835 and Z3146
are an order of magnitude or more larger than those of the rest of the
sample.  A2390 is an intermediate case in that its 15 $\mu$m
luminosity is only slightly excessive ($\sim$ 5 $\sigma$), but the 70
$\mu$m photometry clearly shows that it has a prominent far-infrared
peak in its SED.

These three infrared-brightest BCGs are located in the three most
X-ray luminous clusters in our sample, with soft X-ray (0.1--2.4 keV)
luminosities of $16-21 \times 10^{44}$ erg s$^{-1}$.  These three
clusters also have short gas cooling times of 0.6--1.9 Gyr based on
the latest {\em Chandra} results \citep{Bauer05} as shown in
Table~\ref{flux}.

\subsection{Radio Emission}

\begin{figure}
  \epsscale{0.8}

  \hspace*{-1.5cm}\plotone{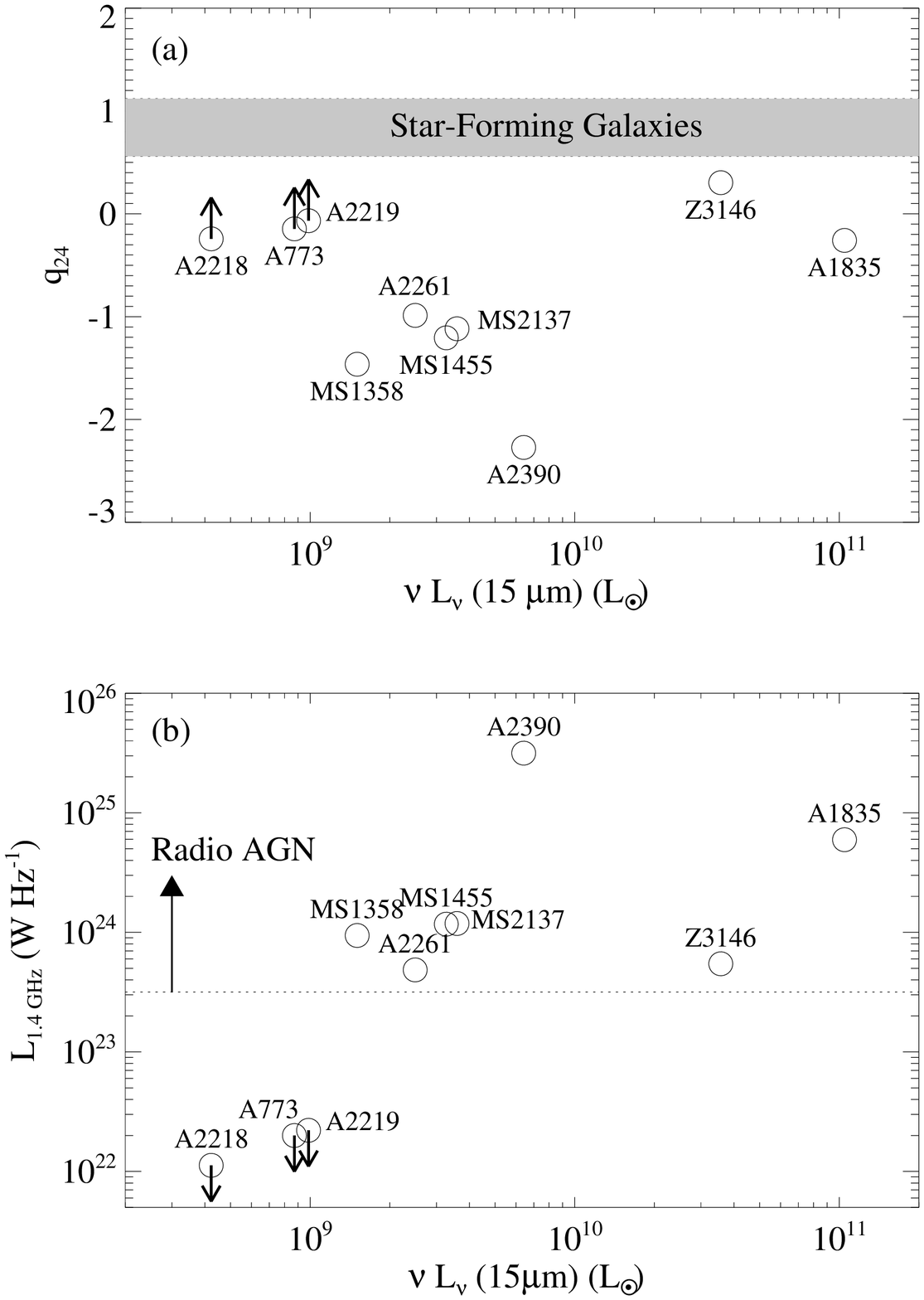}

  \caption{(a) The $q_{24}$ parameter ($q_{24}=\log (S_{24 \mu{\rm
  m}}/S_{20 {\rm cm}}$)) of the BCGs plotted against their 15 $\mu$m
  monochromatic luminosities.  The shaded horizontal band shows the 1
  $\sigma$ range of the $q_{24}$ parameter observed for the 24
  $\mu$m-detected star-forming galaxies \citep{Appleton04}; (b) The
  1.4 GHz radio luminosities ($L_{\nu}$) of the BCGs plotted against
  their 15 $\mu$m monochromatic luminosities.  The radio luminosities
  were calculated as $L=4 \pi D_{L}^{2} S_{1.4GHz}$ (i.e., applying a
  K correction assuming a radio continuum shape of $f_{\nu} \propto
  \nu^{-1}$.)
  \label{radio}}

\end{figure}

Ten of the eleven BCGs (i.e., except for AC114) have been observed in
the radio (Table~\ref{spitzer}).  Figure~\ref{radio}a plots the ratio
of the flux densities at 24 $\mu$m and 20 cm parameterized as the
$q_{24}$ parameter ($q_{24}=\log (S_{24 \mu{\rm m}}/S_{20 {\rm cm}}$))
\citep{Appleton04}.  The $q_{24}$ values of star-forming galaxies fall
in a well-defined range ($0.84 \pm 0.28$) as a result of the
radio-infrared luminosity correlation \citep{Appleton04}, and
therefore this parameter allows us to detect the existence of excess
radio emission likely produced by a radio AGN.  Figure~\ref{radio}a
shows that seven of the ten radio-observed BCGs are clearly out of the
range for star-forming galaxies, having too much 20 cm radio flux
density for their 24 $\mu$m flux density.  The remaining three (those
in A773, A2218, and A2219) are not detected in the radio with
inconclusive lower limits on $q_{24}$.

Based on the 20 cm luminosities, this excessive radio emission in the
BCGs is likely to originate from radio AGNs.  Figure~\ref{radio}b
 shows that the 20 cm luminosities of these BCGs are mostly above
$10^{23.5}$ W Hz$^{-1}$, which is usually the dividing line between
AGNs and star-forming galaxies \citep[e.g.,][]{Yun01}.  All seven BCGs
outside the range of star-forming galaxies in Figure~\ref{radio}a fall
in the radio AGN category in Figure~\ref{radio}b.  The prevalence of
radio AGNs in the BCGs means that there may be a significant AGN
contribution to the observed infrared luminosities of these galaxies.

The three BCGs with a low 20cm luminosity ($\ll 10^{23.5}$ W
Hz$^{-1}$), A773, A2218, and A2219, are all located in non-cooling
flow clusters with a radiative gas cooling time of longer than 10 Gyr
as categorized by \citet{Allen98} (Table~\ref{flux}).  They have the
longest cooling times (24.8, 30.9, and 29.1 Gyr) of the ten
radio-observed clusters based on the latest estimates by
\citet{Bauer05} (Table~\ref{flux}).  Such a trend is consistent with
the earlier finding that there exists a strong coupling between
cluster cooling cores and radio emission associated with the BCGs
\citep[e.g.,][]{Burns90}, which may suggest that appreciable cooling
of the intracluster gas in the cluster core tends to trigger a radio
AGN in the BCG.  The three clusters with no radio detection are also
known to show bi/tri-modal dark matter distributions based on the
lensing analysis by \citet{Smith05}.  Such complex cluster mass
distributions, together with the disturbed X-ray morphologies, may be
related to the long gas cooling times.

\subsection{Properties of the Three Infrared-Brightest BCGs}

Here, we examine in more detail the properties of the three
infrared-brightest BCGs in the sample, those in A1835, Z3146, and
A2390.  Various properties of these three BCGs are summarized in
Table~\ref{three_bright}.  

\begin{figure}
  \epsscale{0.55}

  \hspace*{-0.5cm}\plotone{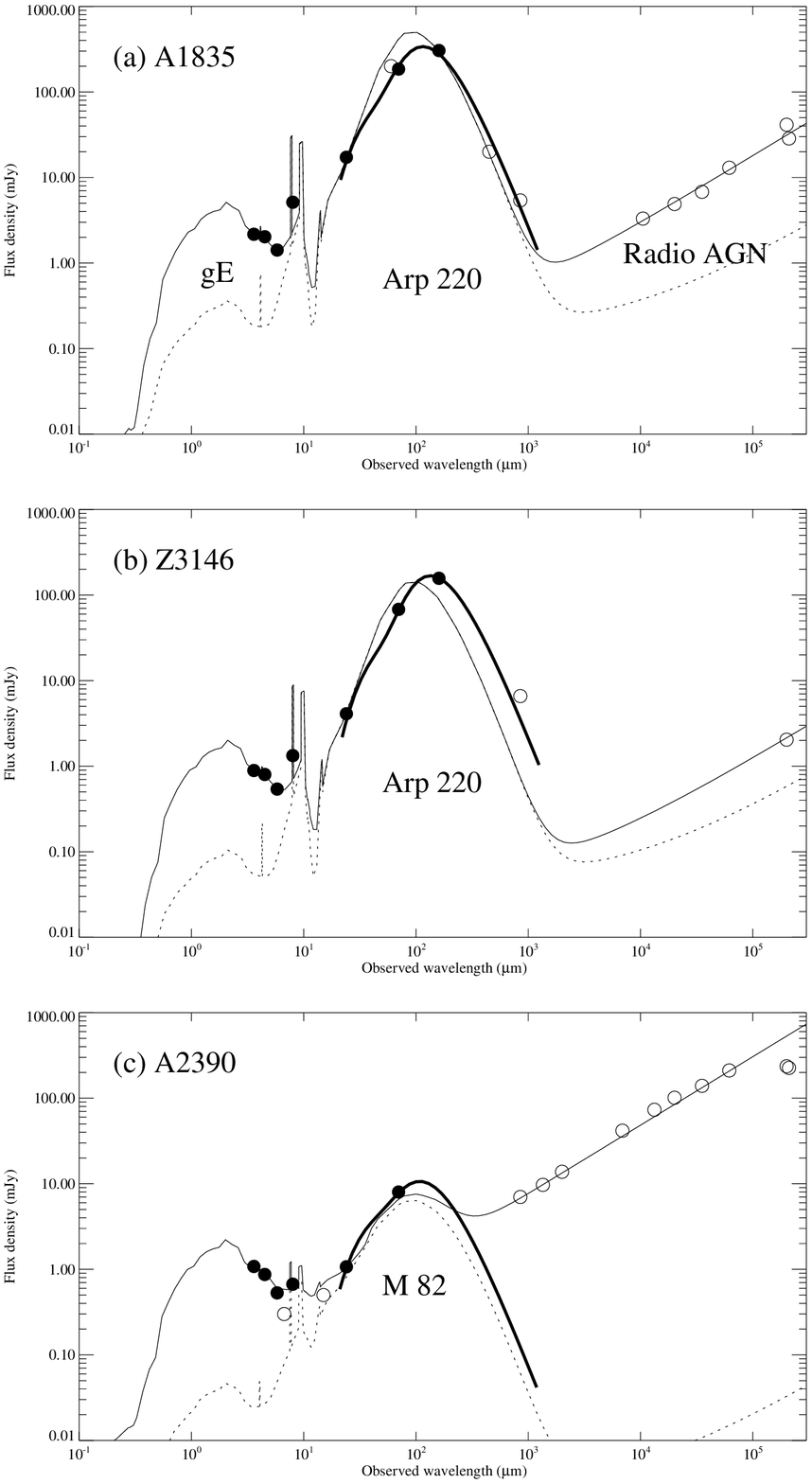}

  \caption{SEDs of the three infrared-brightest BCGs.  The solid
  circles indicate the {\em Spitzer} measurements while the open
  circles indicate the measurements in the literature (the error bars are
  comparable or smaller than the symbols). The dotted line shows the
  SEDs of Arp 220 (a and b) and M~82 (c) \citep{Silva98}.  The thin
  solid line shows the three-component SED model while the thick solid
  line shows the two-component modified black-body SED model to fit
  the observed far-infrared/submm SED (see the text for details).  The
  flux density measurements of the BCGs in A1835 and A2390 were taken
  from \citet{Edge99}.  The visual/near-infrared measurements were
  omitted because of the unknown aperture corrections, but they show a
  sharp drop toward shorter wavelengths typical of the SED of a giant
  elliptical.  For A2390, we also excluded the low signal-to-noise
  IRAS measurements.  The measurements for Z3146 were taken from
  \citet{Chapman02} (850 $\mu$m) and the FIRST survey (20 cm).
  \label{a1835_z3146_sed}}

\end{figure}

Figure~\ref{a1835_z3146_sed} shows the SEDs of the three BCGs.  A
number of photometric measurements in the literature are shown
together with the Spitzer measurements.  We modeled their behavior
with a combination of the following three SED components: a giant
elliptical (gE) SED, an infrared-luminous galaxy SED (Arp~220/M~82),
and a power-law radio spectrum simulating the radio emission from a
radio AGN.  The gE SED was taken from the SED template library of the
Hyper-z photometric redshift code \citep{Bolzonella00}.  For
infrared-luminous galaxy SEDs, we used those of Arp~220 and M~82 taken
from \citet{Silva98}.  The SEDs of \citet{Silva98} do not have the
precision to reproduce individual spectral features (e.g., PAH
features), but provide accurate broad-band SEDs from ultraviolet to
radio.

The results are shown in Figure~\ref{a1835_z3146_sed}.  As shown in
panel (a), when the three components are combined, the gE SED
dominates at $\lambda_{rest} \la 5 \mu$m , the radio AGN SED dominates
at $\lambda_{rest} \ga 1 $mm, and the infrared-luminous galaxy SED
dominates between 5 $\mu$m and 1 mm.  The scalings of the gE SED and
radio AGN SED were determined by the IRAC and radio measurements,
respectively; the scaling of the infrared-luminous galaxy SED was set
by the observed 24 $\mu$m flux density alone, and no effort was made
to fit the far-infrared/submm measurements.  As shown in the figure,
the BCGs in A1835 and Z3146 have a SED steeply rising from 24 to
70/160 $\mu$m, which requires an Arp~220-like SED to fit without
violating the observed gE SED in the restframe visual/near-infrared.
The BCG in A2390, on the other hand, has an SED significantly flatter
between 24 and 70 $\mu$m, and in this case an M~82-like SED is more
appropriate.

Figure~\ref{a1835_z3146_sed} shows that the three-component model
works reasonably well to reproduce the observed SEDs of these three
BCGs (the thin solid lines).  However, the figure also shows that the
use of the template SEDs from a small number of well-known galaxies
(Arp~220, M~82) is clearly too simplistic, and does not work in
detail.  For example, the far-infrared SED peak at $\sim 100
\mu$m seen with the BCGs in A1835 and Z3146 is clearly longward of
that seen with Arp~220.

To determine the infrared luminosities of these BCGs accurately, we
also fitted the far-infrared/submillimeter SEDs with a two-component
modified black-body model\footnote{Each modified black-body model was
represented by the following analytical expression: $F_{\nu} =
B_{\nu}(T_{d})(1-e^{-(\nu/\nu_{0})^{\beta}})$, where $B_{\nu}(T_{d})$
is the Planck function with a dust temperature $T_{d}$, $\nu_{0}$ is
the frequency below which the thermal dust emission becomes optically
thin, and $\beta$ is the dust emissivity power-law index.  A good fit
was obtained for the BCG in A1835 with $T_{d} = 40$ \& 110 K, $\nu_{0}
= 3$ THz (i.e., 100 $\mu$m), and $\beta=1.5$.  The same dust
temperatures were adopted for the BCG in A2390, which does not have
any far-infrared/submillimeter measurements to constrain the SED,
while dust temperatures of 35 and 100 K were used for the BCG in
Z3146.} (the thick solid lines).  These fits produced infrared
luminosities of $7.3 \times 10^{11}$, $4.1 \times 10^{11}$, and $0.3
\times 10^{11}$ L$_{\sun}$ for the BCGs in A1835, Z3146, and A2390,
respectively.  This also means that the BCGs in A1835 and Z3146 can be
classified as LIRGs.  Although Arp~220 is technically a ULIRG, its
SED fits those of the BCGs in A1835 and Z3146 presumably because the
infrared luminosity of Arp~220 ($1.6\times 10^{12} L_{\sun}$ by
\citet{Sanders03}) is barely above the ULIRG threshold.  The infrared
luminosity of the BCG in A2390 is also comparable to that of M~82
($0.6 \times 10^{11} L_{\sun}$ by \citet{Sanders03}).

The models in Figure~\ref{a1835_z3146_sed} clearly show that the radio
continuum flux is well above that expected from the Arp~220 SED.  This
is why these BCGs have small $q_{24}$ values outside the range for
star-forming galaxies (Figure~\ref{radio}a), indicative of a radio
AGN.  However, the radio SEDs of the BCGs in both A1835 and A2390 show
that the slope of the radio spectrum is steep ($\alpha \sim 0.8$,
where $F_{\nu} \propto \nu^{-\alpha}$), and as a result the infrared
luminosity is dominated by thermal dust emission in both BCGs, as
found in general for steep spectrum radio galaxies \citep{Shi05}.  In
the case of the BCG in Z3146, we do not have enough radio measurements
to constrain the radio spectrum, but it is already clear from the SED
in Figure~\ref{a1835_z3146_sed} that it also has a large far-infrared
peak.

Despite the presence of a radio AGN, all the BCGs show infrared SEDs
typical of star-forming galaxies.  In fact, none of the BCGs shows a
power-law infrared SED typical of an AGN-dominated galaxy, and all of
them show an excess of 8 $\mu$m emission, which is likely due to the
6.2 $\mu$m PAH feature in the passband at these redshifts.  In the
case of the BCG in Z3146, the existence of a strong 6.2 $\mu$m feature
has been confirmed spectroscopically with a {\em Spitzer}/IRS
observation (E.~Egami et al.\ 2006, in preparation), which indicates that
star formation is a major contributor to the infrared luminosity.

A number of arguments indicate that the infrared luminosities are
produced by star formation.  From the infrared luminosities
themselves, we derive star formation rates (SFRs) of 125, 70, and 5
M$_{\sun}$ yr$^{-1}$ for the BCGs in A1835, Z3146, and A2390,
respectively, using the relation of \citet{Kennicutt98a}.  Similar
star formation rates are derived from the extinction-corrected
H$\alpha$ luminosities measured by \citet{Crawford99} with the
resultant values of 40, 47, and 5 M$_{\sun}$ yr$^{-1}$
(Table~\ref{three_bright}).  The H$\alpha$-derived star formation
rates are often lower than those derived from the infrared
luminosities (e.g., a factor of three lower with the BCG in A1835)
probably due to the finite slit width (1\farcs2--1\farcs3) of the
optical long-slit spectroscopy \citep{Crawford99} and/or internal dust
extinction.  The infrared luminosities from the restframe 15 $\mu$m
luminosities using the relation of \citet{Elbaz02} for star-forming
galaxies are $9.5\times10^{11}$, $3.3\times10^{11}$, and $0.6 \times
10^{11}$ L$_{\sun}$, similar to the values directly measured above.

One major difference between the SEDs of these BCGs and those of
infrared-luminous galaxies is that the former are much more luminous
in the restframe visual and near-infrared.  At $\lambda_{rest}< 5
\mu$m, the massive old stellar population in the BCGs completely
overwhelms the light associated with the star-forming population.  In
fact, the total luminosity emitted in the visual/near-infrared from
the massive old stellar population is comparable to that emitted by
dust in the infrared.  This explains why it has been difficult to
detect the starbursting component in these BCGs.  A similar object
(i.e., a giant elliptical containing a starbursting population) has
also been found in the field by \citet{Krause03}.

\begin{figure}
  \epsscale{0.7}

  \hspace*{-1cm}\plotone{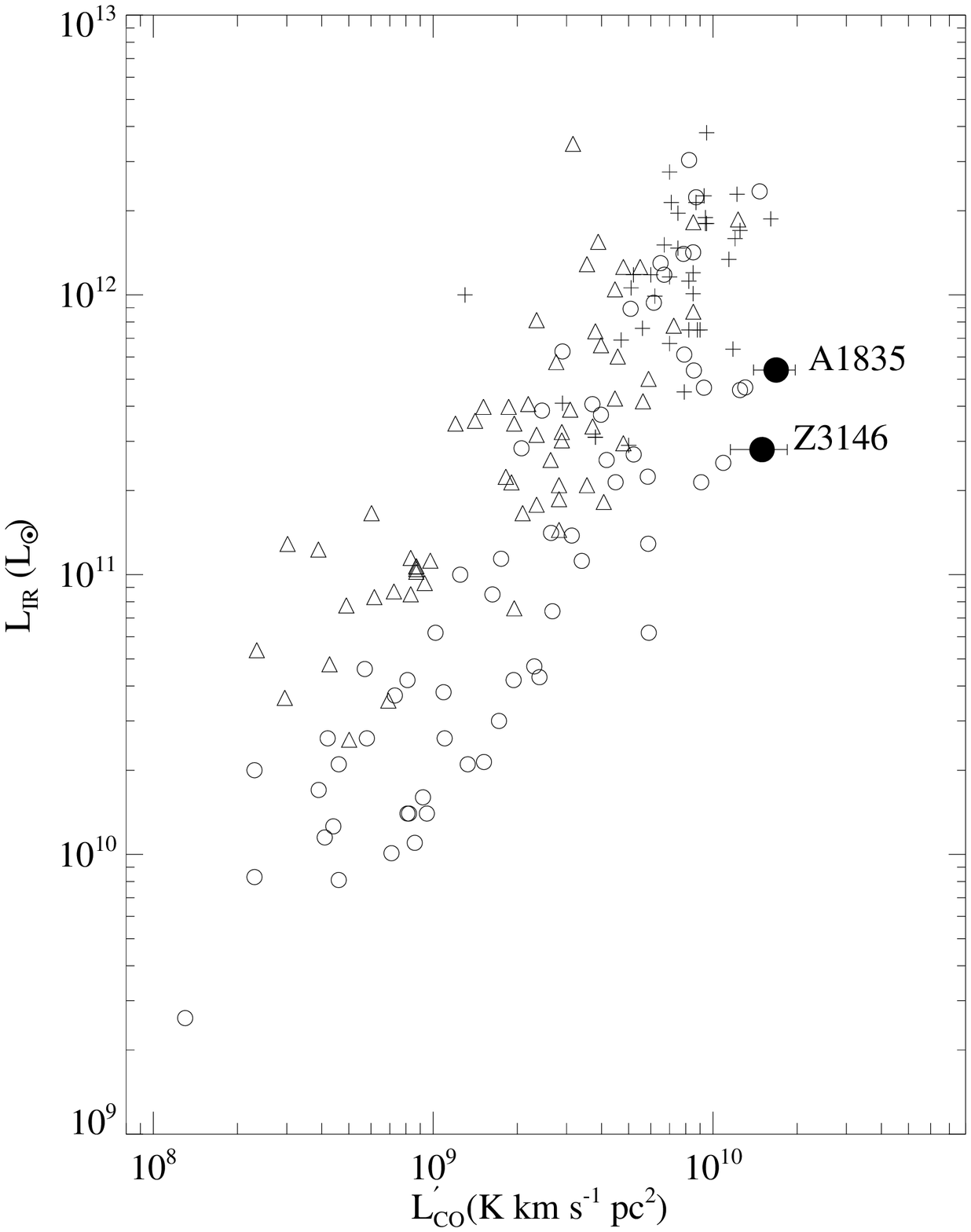}

  \caption{The infrared luminosity and CO(1-0) line luminosity of the
  BCGs in A1835 and Z3146 (solid circles) are compared with those of
  LIRGs and ULIRGs.  See the notes of Table~\ref{three_bright} for the
  definition of the CO line luminosity, $L^{\prime}_{CO}$.  The
  LIRG/ULIRG points shown are from \citet{Sanders91} (triangles),
  \citet{Solomon97} (crosses), and \citet{Gao04a} (open circles).  The
  infrared luminosity and CO line luminosity of the two BCGs listed in
  Table~\ref{three_bright} were rescaled to the values with $H_{0} =
  75$ km s$^{-1}$ Mpc$^{-1}$ and $q_{0}=0.5$ so that the direct
  comparison with the LIRG/ULIRG sample is possible.
  \label{co}}
\end{figure}

The two infrared-brightest BCGs, those in A1835 and Z3146, show
another interesting property: exceptional strength of the CO (1-0)
emission line for their infrared luminosities.  Figure~\ref{co} shows
a broad correlation that holds between the infrared luminosities and
CO luminosities of LIRGs and ULIRGs.  Plotted in the same figure, the
BCGs in A1835 and Z3146 hint at the possibility that they are
overluminous in CO luminosity when compared with LIRGs and ULIRGs with
similar infrared luminosities.  This may mean that these two
BCGs contain an unusually large amount of cold molecular gas.

\subsection{Implications for the Cluster Cooling Flows}

Figure~\ref{plot_24_tcool} plots the 15 $\mu$m monochromatic
luminosities of the BCGs as a function of the radiative cooling time
of the intracluster gas calculated by \citet{Bauer05}.  Over the whole
sample, there seems to be a trend that BCGs in clusters with shorter
gas cooling times (and therefore larger X-ray luminosities in general)
have larger infrared luminosities.  Such a correlation would indicate
that the infrared activity of a single galaxy at the cluster center is
somehow connected to the property of a cluster core as a whole.

\begin{figure}
  \hspace*{1cm}\includegraphics[angle=90,scale=0.6]{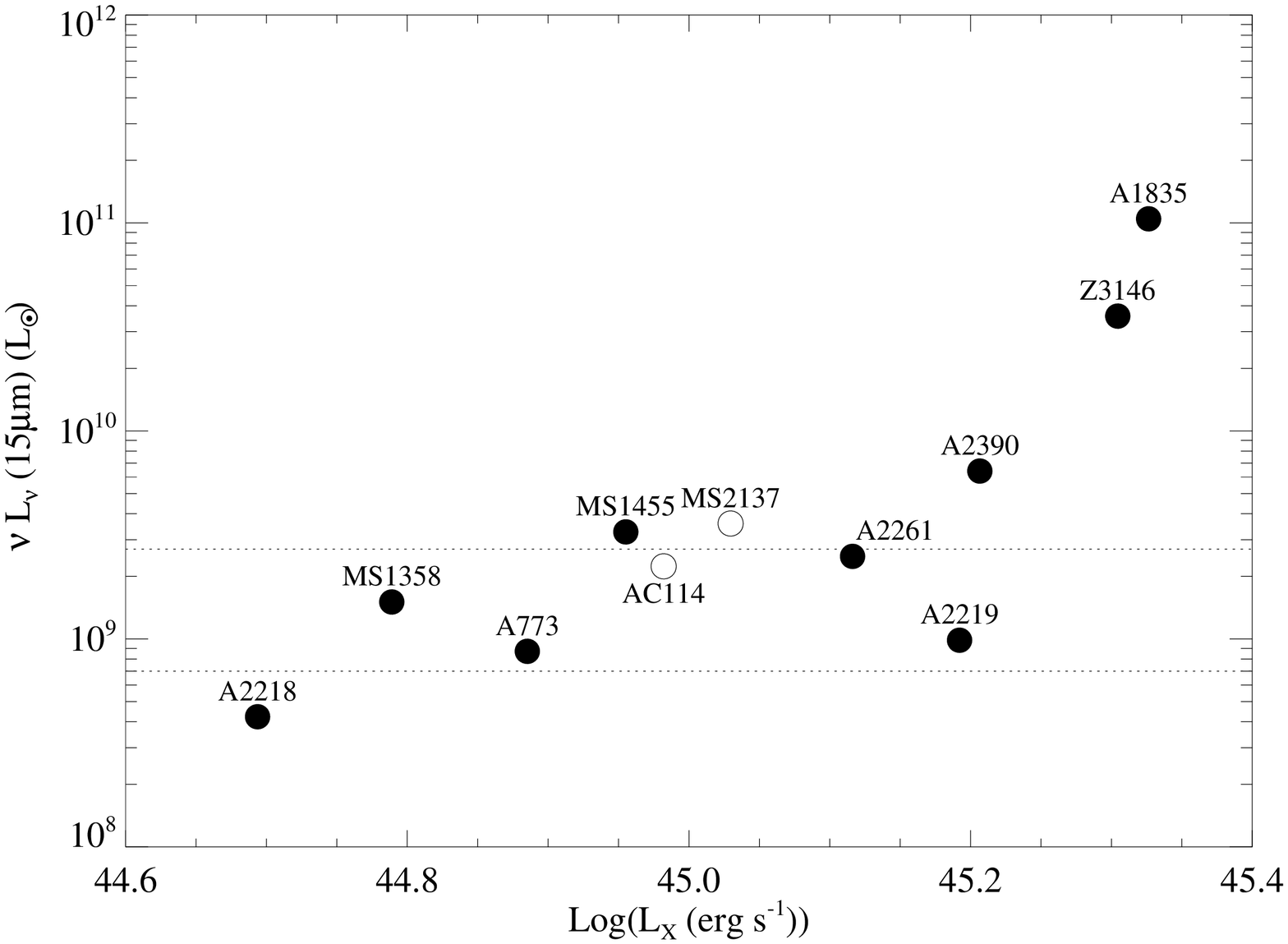}

  \caption{The 15 $\mu$m monochromatic luminosities ($\nu L_{\nu}$) of
  the BCGs are plotted against the cluster soft X-ray (0.1--2.4 keV)
  luminosities from the NORAS survey \citep{Boehringer00}.  For the
  two southern clusters MS2137 and AC114 (unfilled circles), we used
  the Einstein 0.3--3.5 keV \citep{Gioia94} and ASCA 2--10 keV
  \citep{Allen00} X-ray luminosities, respectively.  The latter was
  divided by a factor of 1.25 to account for the systematic offset
  between the ROSAT soft X-ray luminosities and ASCA hard X-ray
  lumninosities.  The horizontal dotted lines show the 1 $\sigma$
  range of the 15 $\mu$m monochromatic luminosities observed for local
  elliptical galaxies \citep{Ferrari02}, which is $\nu L_{\nu}[15
  \mu{\rm m}] = (1.7 \pm 1) \times 10^{9} L_{\sun}$ (Note: these 15
  $\mu$m luminosities are not those tabulated by \citet{Ferrari02},
  which do not contain the stellar contribution, but those directly
  calculated from their flux density measurements.)
  \label{plot_24_x}}

\end{figure}

\begin{figure}
  \hspace*{1cm}\includegraphics[angle=90,scale=0.6]{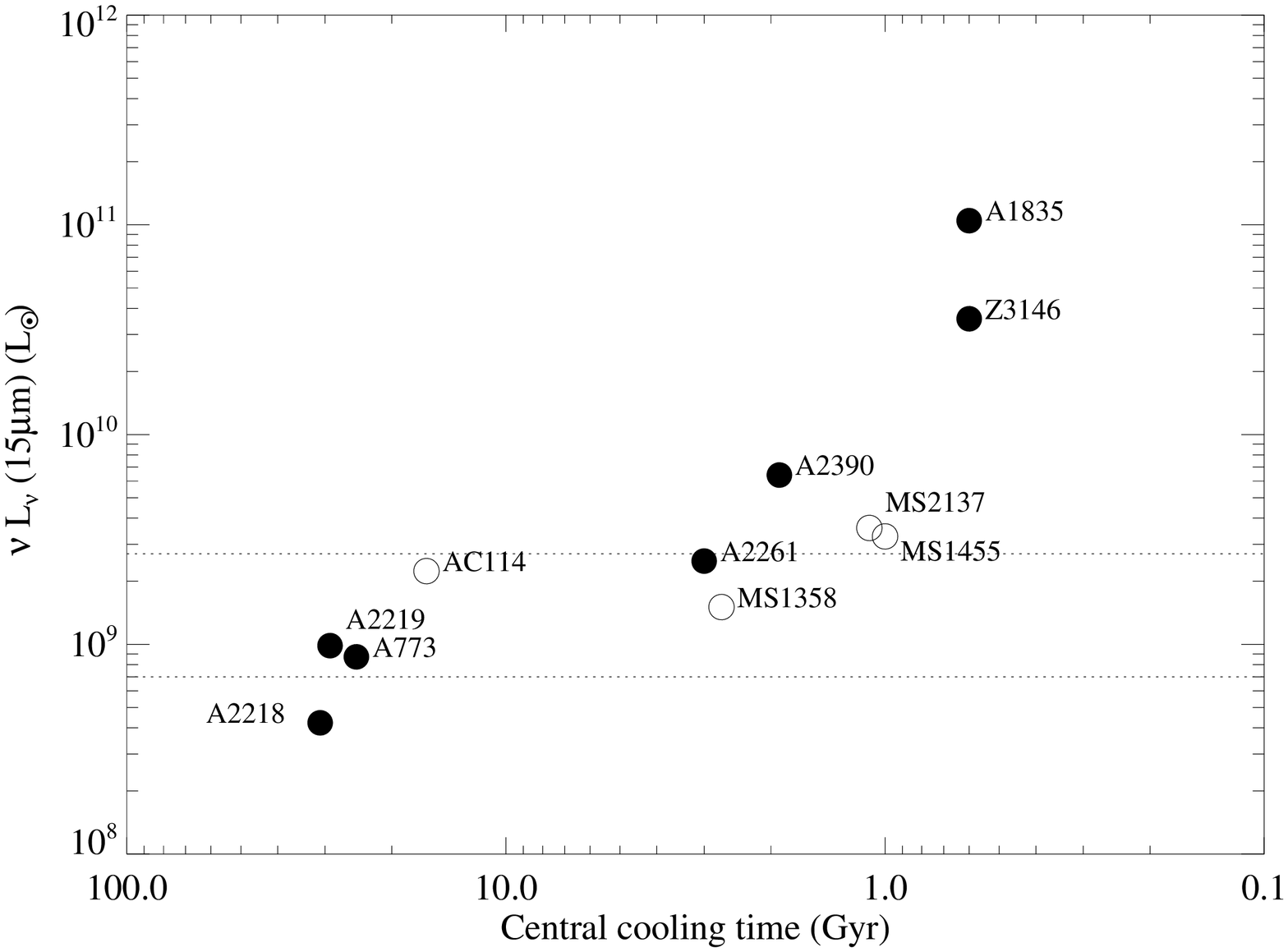}

  \caption{The 15 $\mu$m monochromatic luminosities of the BCGs are
  plotted against the radiative cooling times of cluster gas derived
  from {\em Chandra} observations by \citet{Bauer05} (solid circles)
  and from {\em ROSAT} observations by \citet{Allen00} (open circles).
  The horizontal dotted lines show the 1 $\sigma$ range of the 15
  $\mu$m monochromatic luminosities observed for local ellipticals
  shown in Figure~\ref{plot_24_x}.
  \label{plot_24_tcool}}
\end{figure}

Note, however, that a large X-ray luminosity does not always translate
into a short cooling time.  For example, A2219 has a large X-ray
luminosity but a long cooling time.  In fact, A2219 is a peculiar
cluster in that it highly deviates from the cluster X-ray
luminosity--temperature ($L_{X}$ vs.\ $T$) relation and
mass--temperature ($M$ vs.\ $T$) relation \citep{Smith05}.  This
cluster may have experienced a core-penetrating merger recently
\citep{Smail95}, and such a process might have boosted the X-ray
temperature and luminosity \citep{Smith05}.  In any case, the fact
that A2219 deviates from the overall trend in Figure~\ref{plot_24_x}
but not in Figure~\ref{plot_24_tcool} suggests that the correlation
between the 15 $\mu$m luminosity and gas cooling time is more
fundamental than that between the 15 $\mu$m luminosity and X-ray
luminosity.

In general, the increased infrared luminosities observed in local
galaxies are thought to be episodic.  For example, transient
galaxy-galaxy interactions are believed to boost the infrared
luminosity by triggering vigorous star-formation and/or fueling an
AGN.  However, the process that triggered the increased infrared
activity in the BCGs in A1835 , Z3146, and A2390 is difficult to
explain in terms of similar processes since in this case the infrared
luminosity of a BCG would likely increase independent of the
properties of the cluster in which it resides.  In other words, in
this scenario there is nothing that would prevent a BCG in a low X-ray
luminosity and long cooling time cluster from becoming
infrared-luminous.  This would introduce a significant scatter in the
correlations of the BCG mid-infrared luminosity with the cluster X-ray
luminosity (Figure~\ref{plot_24_x}) and with the cluster radiative gas
cooling time (Figure~\ref{plot_24_tcool}).  The fact that we do not
see significantly deviant points from these trends suggests that the
triggering mechanism is not a stochastic process but rather a process
that is controlled by the properties of the cluster cores as a whole.

An attractive possibility is to associate the observed infrared
luminosity with the fate of the cooling cluster gas.  In this case, only
those BCGs in clusters with a short gas cooling time could become
infrared-luminous because only these clusters could channel the gas
into the BCGs and form stars.  This hypothesis can be tested by
comparing the infrared-derived SFRs with the X-ray-derived mass
deposition rates.

Table~\ref{mdot} lists various determinations of the mass deposition
rates for A1835, Z3146, and A2390 based on the X-ray spectral
analysis.  Together with the previously published values, we also list
our own estimates based on the reanalysis of the {\em Chandra} data.
We calculated the mass deposition rates at the cluster center within a
radius of 13\arcsec\ ($\sim$ 50 kpc) from the BCGs (i.e., the size of
the 24 $\mu$m photometry beam) as well as for the entire clusters ($r
< 500-700$ kpc).  The central mass deposition rates are more relevant
for comparison with the star formation rates in the BCGs, and can be
compared with other published {\em Chandra}-derived mass deposition
rates, which were also calculated within comparable radii
(Table~\ref{mdot}).  We did not list the mass deposition rates derived
from the {\em ROSAT} and {\em ASCA} data because they are believed to
be overestimated by almost an order of magnitude in many cases.  The
large mass deposition rates resulting from the {\em ROSAT} and {\em
ASCA} data, exceeding 1000 M$_{\sun}$ yr$^{-1}$ in extreme cases, are
thought to originate from the existence of cool ambient gas in the
cluster core, which, when spatially unresolved, tends to be included
into the cooling-flow calculation and inflates the mass deposition
rates \citep{Allen01}.  The difference between the central ($r<50$
kpc) and total ($r<500-700$ kpc) mass deposition rates listed in
Table~\ref{mdot} reflects the size of this effect.

Table~\ref{mdot} shows that the various estimates of the mass
depostion rates are not always consistent.  In the case of A1835, for
example, the value of 130--200 M$_{\sun}$ yr$^{-1}$ ($r< 24$ kpc)
derived by \citet{Schmidt01} is significantly larger than the other
estimates.  In fact, \citet{Voigt04} point out that the existing
Chandra data of A1835 at $r<100$ kpc is statistically consistent with
the absence of the cooling-flow component, and our reanalysis of the
same data is in line with this result.  In the case of A2390, the
various estimates are broadly consistent with each other, especially if
we account for the effect mentioned above that would inflate the mass
deposition rate as we increase the region to extract the X-ray
spectrum.  In the case of Z3146, we do not have any other estimates to
compare at this point, but this cluster provides the most convincing
evidence for having a significant mass deposition rate.  The
differences between the various estimates probably reflect the level
of uncertainty inherent in such analyses, and some part of the
differences may be attributed to the improvements in the {\em
Chandra}/ACIS calibration since the early years of the mission.

The infrared-derived SFRs for the BCGs in A1835, Z3146, and A2390 are
125, 70, and 5 M$_{\sun}$ yr$^{-1}$, respectively, while the
X-ray-derived mass deposition rates at the cluster center are $<20$,
290, and 110 M$_{\sun}$ yr$^{-1}$, respectively, based on our own
estimates.  This indicates that the correspondence between the
infrared-derived SFRs and X-ray-derived mass deposition rates is
currently suggestive at best.  Z3146 is the only case where the two
quantities are broadly consistent (within a factor of several).  A2390
shows a large difference (a factor of 20), which may be related to the
strong radio AGN detected in the BCG of this cluster.  In the case of
A1835, the difference seems at lease a factor of six although we are
in need of deeper {\em Chandra} data to better constrain the mass
deposition rate and reconcile the conflicting estimates for the mass
deposition rate.

Given the complex nature of star formation processes and the limited
accuracy with the determination of the physical quantities such as
star formation rates and mass deposition rates, it would not be
surprising that the infrared-derived SFRs and X-ray-derived mass
deposition rates differ by a factor of several or more even if there
is a causal connection between the two.  For example, the well-known
Schmidt law (i.e., the correlation between the surface densities of
gas and star formation rate seen in external galaxies) has a 1
$\sigma$ scatter of a factor of 2, and individual galaxies deviate by
as much as a factor of 7 \citep{Kennicutt98b}.  Therefore, at least in
Z3146 (and possibly in A1835), it is possible to interpret the
observed SFRs as a result of the cooling cluster gas accreting onto
the BCGs and triggering star formation (i.e., the cooling-flow
interpretation).  Such a scenario is also consistent with the the
exceptionally strong CO line emission observed with these two BCGs
since cold molecular gas is seen to be accreting to the BCG in one of
the low-redshift cooling-flow clusters \citep{Salome04}.

\section{Conclusions}

With the {\em Spitzer}/IRAC (3.6--8.0 $\mu$m) and {\em Spitzer}/MIPS
(24--160 $\mu$m) imaging data, we studied the infrared properties of the
brightest cluster galaxies (BCGs) located in the cores of
X-ray-luminous clusters.  The main results are as follows:

\begin{itemize}
  \item The majority of BCGs are not infrared-luminous.  Out of the 11
  BCGs studied here, only three (those in A1835, Z3146, and A2390)
  have a strong far-infrared SED peak, and only two (those in A1835
  and Z3146) can be classified as LIRGs ($>10^{11}$ L$_{\sun}$).

  \item There seems to be a trend that clusters with larger X-ray
  luminosities and shorter gas cooling times harbor more
  infrared-luminous BCGs.  The three infrared-brightest BCGs (those in
  A1835, Z3146, and A2390) were found in clusters with the
  largest X-ray luminosities and shortest radiative gas cooling times
  in the sample.

  \item The infrared luminosities of the three infrared-brightest BCGs
  are likely to be powered by star formation judged from the infrared
  SED signatures.  The infrared-derived SFRs are comparable to the
  extinction-corrected H$\alpha$-derived SFRs.

  \item Radio AGNs are prevalent in the BCGs.  Seven out of the ten
  radio-observed BCGs studied here contain a luminous radio source
  ($>10^{23.5}$ W Hz$^{-1}$ at 20 cm), and show a low 24 $\mu$m/20 cm
  flux density ratio, both indicative of a radio AGN.

  \item The infrared-derived SFR of the BCG and the X-ray-derived mass
  deposition rate in the parent cluster are broadly consistent in
  Z3146 (and possibly in A1835), suggesting the interpretation that
  the observed star formation is triggered by the cooling cluster gas
  accreting onto the BCGs except when there is a strong radio AGN as
  is the case in A2390.

\end{itemize}

\acknowledgments

This work is based on observations made with the {\em Spitzer} Space
Telescope, which is operated by the Jet Propulsion Laboratory,
California Institute of Technology under a contract with NASA
(contract number \#1407).  Support for this work was provided by NASA
through an award issued by JPL/Caltech (contract number \#960785,
\#1255094).  GPS acknowledges financial support from Caltech and a
Royal Society University Research Fellowship.

\appendix

\section{Reanalysis of the {\em Chandra} X-ray data\label{chandra}}

The {\em Chandra} data were re-analyzed using CIAO 3.2 and the latest
calibration files available in CALDB 3.1.  Each individual dataset was
reprocessed and screened to remove periods of strong background flares
in the standard manner.  Matching background event files were created
for each dataset from the standard ACIS blank-sky event files
following the procedure described in \cite{Vikhlinin05}.  

For a given cluster, we have calculated the total gas cooling rate for
the entire cluster as well as the cooling rate in the immediate
surrounding of the BCG (within a radius of 13\arcsec, which is the
size of the 24 $\mu$m photometry beam).  The total cooling rates were
calculated based on fits to the integrated spectrum inside a maximum
outer radius. In practice, this radius was determined from the radial
surface brightness and defined as the point at which the surface
brightness profile drops below the background level.  For A1835 and
A2390, the individual datasets were combined to form a mosaic and the
radial surface brightness determined for this composite. With this
definition, the corresponding maximum radii are $125 \arcsec$ ($\sim
490$ kpc), $125 \arcsec$ ($\sim 460$ kpc), and $160 \arcsec$ ($\sim
700$ kpc) for A1835, A2390, and ZW3146, respectively.

To calculate the gas cooling rates, spectra in a given region were
extracted in the 0.5--7.0 keV energy range along with corresponding
background spectra from the blank-sky event files mentioned
previously. Counts weighted detector response (RMFs) and effective
area (ARFs) files were created for the extraction regions using the
CIAO tools {\tt mkacisrmf} and {\tt mkwarf}.  The resulting spectra
were then compared to various standard spectral models using the XSPEC
model library available in the ISIS spectral fitting package
\citep{Houck00}. The fitted spectral model was defined to consist of
foreground Galactic absorption, a single temperature thermal plasma
(MEKAL) due to emission from the hot, outer cluster, and an isobaric
cooling flow component (MKCFLOW). Fits were performed with the
Galactic column fixed to the standard value as well as allowing it to
varying. For the two clusters with multiple observations, individual
spectra were extracted for each dataset and the model was fit to the
component spectra simultaneously. For A1835 and ZW3146, the fits with
and without varying Galactic columns gave comparable $\chi^2$ values
and the cooling rates quoted in Table~\ref{mdot} are for the
fixed column fits. Fits for A2390 showed a marked improvement in
$\chi^2$ if the Galactic column was increased by $\sim$30\% and the
result from this fit is quoted in Table~\ref{mdot}.

\clearpage

\begin{deluxetable}{lccccccccccl}
\tablecaption{Target Clusters and Their X-ray Properties \label{flux}}
\tablewidth{0pt} 
\tabletypesize{\scriptsize}
\rotate
\tablehead{ 
\colhead{Name\tablenotemark{a}} & 
\colhead{RA\tablenotemark{b}} & \colhead{DEC\tablenotemark{b}} &
\colhead{$z$\tablenotemark{c}} & 
\multicolumn{4}{c}{X-ray Luminosity} &
\colhead{} &
\multicolumn{2}{c}{Gas Cooling Time} &
\colhead{Type\tablenotemark{j}} \\
\cline{5-8} \cline{10-11} \\
\colhead{} & \colhead{} & \colhead{} & \colhead{} & 
\colhead{ROSAT (NORAS)\tablenotemark{d}} &
\colhead{ROSAT (BCS)\tablenotemark{e}} &
\colhead{ASCA\tablenotemark{f}} &
\colhead{Einstein (EMSS)\tablenotemark{g}} & 
\colhead{} &
\colhead{ROSAT\tablenotemark{h}} &
\colhead{Chandra\tablenotemark{i}} &
\colhead{} \\
\colhead{} & \colhead{} & \colhead{} & \colhead{} &
\colhead{($10^{44}$erg s$^{-1}$)} & 
\colhead{($10^{44}$erg s$^{-1}$)} & 
\colhead{($10^{44}$erg s$^{-1}$)} &  
\colhead{($10^{44}$erg s$^{-1}$)} & 
\colhead{} &
\colhead{(Gyr)} &
\colhead{(Gyr)} & 
\colhead{}
}

\startdata
A773            & 09 17 53.4 & $+$51 43 39 & 0.217 &  7.66   &  8.27   &  9.36 & \nodata & &  9.1 & 24.8    & NCF\\
Z3146           & 10 23 39.6 & $+$04 11 12 & 0.291 & 20.18   & 17.86   & 24.89 & \nodata & &  1.6 &  0.6    & CF\\
MS1358.1$+$6245 & 13 59 50.6 & $+$62 31 05 & 0.327 &  6.14   & \nodata &  7.51 & 7.38    & &  2.7 & \nodata & CF\\
A1835           & 14 01 02.0 & $+$02 52 45 & 0.252 & 21.24   & 25.14   & 29.68 & \nodata & &  1.4 &  0.6    & CF\\
MS1455.0$+$2232 & 14 57 15.1 & $+$22 20 35 & 0.258 &  9.00   &  8.65   & 10.89 & 10.51   & &  1.0 & \nodata & CF\\
A2218           & 16 35 49.3 & $+$66 12 45 & 0.175 &  4.96   &  5.66   &  6.57 & \nodata & &  9.5 & 30.9    & NCF\\
A2219           & 16 40 19.6 & $+$46 42 43 & 0.228 & 15.54   & 13.03   & 24.84 & \nodata & &  8.4 & 29.1    & NCF\\
A2261           & 17 22 27.2 & $+$32 07 58 & 0.224 & 13.11   & 11.57   & 15.21 & \nodata & &  2.8 &  3.0    & CF\\
MS2137.3$-$2353 & 21 40 15.1 & $-$23 39 39 & 0.313 & \nodata & \nodata & 11.68 & 10.73   & &  1.1 & \nodata & CF\\
A2390           & 21 53 36.7 & $+$17 41 45 & 0.233 & 16.13   & \nodata & 26.49 & \nodata & &  3.9 &  1.9    & CF\\
AC114           & 22 58 48.4 & $-$34 48 09 & 0.312 & \nodata & \nodata & 12.01 & \nodata & & 16.2 & \nodata & NCF\\
\enddata

\tablenotetext{a}{The prefix ``A'' denotes the Abell clusters
\citep{Abell89} while the prefix ``MS'' denotes the {\em Einstein}
Extended Medium Sensitivity Survey clusters \citep{Gioia94}.  The
Zwicky cluster Z3146 is often designated as ZwCl1021.0$+$0426 (NED
convention) while the cluster AC114 is also known as the Abell
supplementary southern cluster AS1077 \citep{Abell89}.}

\tablenotetext{b}{These coordinates indicate the positions of the
BCGs. The BCG position in MS2137 was measured in the Palomar Digital
Sky Survey image while other positions were taken from
\citet{Smail01} (A2218), \citet{Stanford02} (MS1358 and AC114), and
\citet{Crawford99} (for the rest), respectively.}

\tablenotetext{c}{All the redshifts were taken from \citet{Allen00}.}

\tablenotetext{d}{0.1--2.4 keV luminosity from NORAS survey \citep{Boehringer00}}
\tablenotetext{e}{0.1--2.4 keV luminosity from BCS survey \citep{Ebeling98}}
\tablenotetext{f}{2--10 keV luminosity from ASCA \citep{Allen00}}
\tablenotetext{g}{0.3--3.5 keV luminosity from the {\em Einstein}
Medium Sensitivity Survey \citep{Gioia94}}

\tablenotetext{h}{Radiative gas cooling times at the cluster center
  from \citet{Allen00} based on the ROSAT observations.}

\tablenotetext{i}{Radiative gas cooling times at the cluster center
  from \citet{Bauer05} based on the {\em Chandra} observations.}

\tablenotetext{j}{Cooling flow (CF)/Non-cooling flow (NCF)
  classifications by \citet{Allen98} (CF: 90\%-conficence
  upper limit on $t_{cool} < 10$ Gyr).  The Chandra cooling times by
  \citet{Bauer05} do not affect the classification at least for these clusters.}

\tablecomments{All the listed values correspond to the cosmological
  parameters ($\Omega_{M}$, $\Omega_{\Lambda}$, $H_{0}$ (km
  s$^{-1}$ Mpc$^{-1}$)) of (0.3, 0.7, 70).  When converted from a
  different choice of the cosmological parameters, the X-ray
  luminosities were scaled with the inverse square of the luminosity
  distance.  Since the gas temperature is independent of the choice of
  the cosmological parameters, the gas cooling time through
  bremsstrahlung emission scales with $n_{e}^{-1} \propto
  \sqrt{\frac{Vp}{L_{X}}}$, where $n_{e}$, $Vp$, $L_{X}$ are the
  electron density, proper volume, and luminosity of the X-ray
  emitting gas.}

\end{deluxetable}

\clearpage

\begin{deluxetable}{lccccccccc}
\tablecaption{Spitzer/VLA Flux Density Measurements of the BCGs \label{spitzer}}
\tablewidth{0pt} 
\tabletypesize{\footnotesize}
\rotate
\tablehead{ 
\colhead{Name} & \colhead{3.6 $\mu$m} & \colhead{4.5 $\mu$m} &
\colhead{5.8 $\mu$m} & 
\colhead{8.0 $\mu$m} &
\colhead{24 $\mu$m}  &
\colhead{70 $\mu$m}  &
\colhead{160 $\mu$m}  &
\multicolumn{2}{c}{1.4 GHz} \\
\colhead{} & \colhead{(mJy)} & \colhead{(mJy)} & \colhead{(mJy)} &
\colhead{(mJy)} & \colhead{(uJy)} & \colhead{(mJy)} & 
\colhead{(mJy)} & \colhead{mJy} & \colhead{ref} 
}

\startdata
A773     & 1.42 & 1.10 & 0.62 & 0.42 &   $98\pm14$   & \nodata    & \nodata    & $<0.14$\tablenotemark{a} & 1 \\
Z3146    & 0.89 & 0.80 & 0.54 & 1.33 &  $4099\pm410$  & $68\pm14$  & $157\pm35$ &  2.04   & 2 \\
MS1358   & 0.69 & 0.56 & 0.32 & 0.24 &    $90\pm13$   & \nodata    & \nodata    &  2.61   & 2 \\
A1835    & 2.17 & 2.03 & 1.42 & 5.14 & $17236\pm1724$ & $185\pm37$ & $305\pm63$ & 31.25   & 2 \\
MS1455   & 1.27 & 1.00 & 0.59 & 0.50 &   $345\pm37$   & \nodata    & \nodata    &  5.62   & 2 \\
A2218    & 1.18 & 0.86 & 0.52 & 0.31 &    $84\pm13$   & \nodata    & \nodata    & $<0.14$\tablenotemark{a} & 1 \\
A2219    & 1.02 & 0.79 & 0.42 & 0.28 &   $125\pm15$   & \nodata    & \nodata    & $<0.14$\tablenotemark{a} & 3 \\
A2261    & 2.15 & 1.66 & 0.90 & 0.65 &   $347\pm36$   & \nodata    & \nodata    &  3.40   & 2 \\
MS2137   & 0.88 & 0.68 & 0.42 & 0.28 &   $294\pm31$   & \nodata    & \nodata    &  3.8    & 4 \\
A2390    & 1.08 & 0.87 & 0.53 & 0.67 &  $1075\pm108$  & $8\pm3$    & \nodata    &  200    & 3 \\
AC114    & 0.95 & 0.79 & 0.48 & 0.30 &   $160\pm18$   & \nodata    & \nodata    & \nodata & \nodata \\
\enddata

\tablenotetext{a}{3 $\sigma$ upper limit ($\sigma \sim 45 \mu$Jy)}

\tablerefs{(1) \citet{Morrison03}; (2) FIRST survey
  \citep{Becker95,White97}. The latest catalog (2003 Apr 11 version)
  is available from \url{http://sundog.stsci.edu}; (3)
  \citet{Rizza03}; (4) NVSS survey \citep{Condon98}}

\tablecomments{All the BCGs are strongly detected in the IRAC bands,
and therefore the IRAC photometric uncertainties are dominated by
those of the absolute calibration.  For the MIPS bands, the 1 $\sigma$
error bars including both the measurement and calibration
uncertainties are explicitly shown because the contribution from the
former is not negligible for weak detections.}

\end{deluxetable}

\clearpage

\begin{deluxetable}{llccc}
\tablecaption{Properties of the Three Infrared-Brightest BCGs\label{three_bright}}

\tabletypesize{\footnotesize}
\tablewidth{0pt}
\tablehead{
\colhead{} & \colhead{} &
\colhead{A1835} &
\colhead{Z3146} &
\colhead{A2390}  
}

\startdata 
$L_{IR}$ \tablenotemark{a}                  & ($10^{11} L_{\sun}$)                & $7.3\pm1.5$           &  $4.1\pm0.8$          &  $0.3\pm0.1$     \\
SFR (IR)\tablenotemark{b}                   & (M$_{\sun}$ yr$^{-1}$)              & $125\pm25$            &  $70\pm14$            &  $5\pm1$ \\
L(H$\alpha$)\tablenotemark{c}               & ($10^{40}$ erg s$^{-1}$)            & $511\pm105$           & $594\pm59$            &  $67\pm18$   \\
SFR (H$\alpha$)\tablenotemark{d}            & (M$_{\sun}$ yr$^{-1}$)              & $40\pm8$              &  $47\pm5$              &  $5\pm1$ \\
$S_{CO} \Delta V$\tablenotemark{e}          & (J km s$^{-1}$)                     & $7.7\pm1.3$           &  $5.2\pm1.2$          &  \nodata     \\
$L_{CO}^{\prime}$\tablenotemark{f}          & ($10^{10}$ K km s$^{-1}$ pc$^{2}$)  & $2.4\pm0.4$           &  $2.2\pm0.5$          &  \nodata     \\
\enddata

\tablenotetext{a}{Infrared luminosity calculated from the
two-component modified black-body SED models shown in
Figure~\ref{a1835_z3146_sed}.}

\tablenotetext{b}{Star formation rate derived as $4.5 \times 10^{-44}
L_{IR}$ (erg s$^{-1}$) \citep{Kennicutt98a}.}

\tablenotetext{c}{Extinction-corrected H$\alpha$ luminosity from
  \citet{Crawford99} rescaled with $\Omega_{M}=0.3$,
  $\Omega_{\Lambda}=0.7$, and $H_{0} = 70$ km s$^{-1}$ Mpc$^{-1}$.}

\tablenotetext{d}{Star formation rate derived as $7.9 \times 10^{-42}
L(H_{\alpha})$(erg s$^{-1}$) \citep{Kennicutt98a}.}

\tablenotetext{e}{Velocity-integrated line flux measured by
\citet{Edge03}}

\tablenotetext{f}{CO(1-0) line luminosity calculated as $
  L^{\prime}_{CO} = 3.25 \times 10^{7} S_{CO} \Delta V \nu_{obs}^{-2}
  D_{L}^{2} (1+z)^{-3}$, where $S_{CO} \Delta V$ is the
  velocity-integrated line flux in Jy km s$^{-1}$, $\nu_{obs}$ is
  observed frequency in GHz, and $D_{L}$ is the luminosity distance in
  Mpc \citep{Solomon97}.}

\tablecomments{For $L_{IR}$ and SFR(IR) (i.e., the first two rows), we
adopted a conservative uncertainty estimate of 20\%, which corresponds
to the absolute calibration uncertainties at 70 and 160 $\mu$m.}

\end{deluxetable}

\clearpage

\begin{deluxetable}{cccccccc}
\tablecaption{Mass Deposition Rates for the Three X-ray-Brightest Clusters\label{mdot}}
\tabletypesize{\footnotesize}
\tablewidth{0pt}

\tablehead{
\colhead{Cluster} &
\multicolumn{5}{c}{\em Chandra} &
\colhead{} &
\colhead{\em XMM} \\
\cline{2-6} \cline{8-8} 
\colhead{} &
\colhead{SAF01\tablenotemark{a}} &
\colhead{AEF01\tablenotemark{a}} &
\colhead{This work\tablenotemark{b}} &
\colhead{VF04\tablenotemark{c}} &
\colhead{This work\tablenotemark{d}} &
\colhead{} &
\colhead{P01} \\
\colhead{} &
\colhead{($<$24 kpc)} &
\colhead{($<$40 kpc)} &
\colhead{($<$50 kpc)} &
\colhead{($<$100 kpc)} &
\colhead{($<$500--700 kpc)} &
\colhead{} &
\colhead{($<120$ kpc)} 
}

\startdata
A1835 & 130--200 & \nodata  & $<20$               & $34^{+43}_{-34}$   & $100^{+100}_{-70}$  & & $<130$  \\
Z3146 & \nodata  & \nodata  & $290^{+100}_{-100}$ & \nodata            & $670^{+160}_{-170}$ & & \nodata  \\
A2390 & \nodata  & 130--190 & $110^{+40}_{-30}$   & $339^{+42}_{-118}$ & $420^{+90}_{-80}$   & & \nodata 
\enddata

\tablenotetext{a}{We adopted the mass deposition rate of 200--300
M$_{\sun}$ yr$^{-1}$ at $r<30$ kpc for A1835 and at $r<50$ kpc for
A2390 ($\Omega_{M}=1$, $\Omega_{\Lambda}=0$, and $H_{0} = 50$ km
s$^{-1}$ Mpc$^{-1}$).}

\tablenotetext{b}{The centeral mass deposition rates derived for the
$r < 13$\arcsec\ region, which is the size of the 24 $\mu$m photometry
beam.  This radius corresponds to 51, 57, and 48 kpc at the distances
of A1835, Z3146, and A2390, respectively.}

\tablenotetext{c}{The deposition rates were calculated inside the
region where the gas cooling time is less then 5 Gyr.  This roughly
corresponds to 100 kpc for A1835 and A2390.}

\tablenotetext{d}{The total mass deposition rates calculated within
the maximum radii of 490, 460, and 700 kpc at the distances of A1835,
A2390, and ZW3146, respectrively.}

\tablecomments{All the numbers are in the unit of M$_{\sun}$
yr$^{-1}$, and the error bars indicate the 90\% confidence limits.
The numbers in the parentheses indicate the radius of the region
inside which the mass deposition rate was calculated.  The mass
deposition rate scales with the inverse square of the luminosity
distance, and all the published values have been rescaled with
$\Omega_{M}=0.3$, $\Omega_{\Lambda}=0.7$, and $H_{0} = 70$ km s$^{-1}$
Mpc$^{-1}$.  The XMM-derived mass deposition rate for A1835 by
\citet{Peterson01} was rescaled with a luminosity distance of 1272
Mpc.}

\tablerefs{SAF01: \citet{Schmidt01}; AEF01: \citet{Allen01}; VF04:
\citet{Voigt04}; P01: \citet{Peterson01}}

\end{deluxetable}

\clearpage
 
\clearpage

\begin{figure}

  \epsscale{0.8}

  \hspace*{-2cm}\plotone{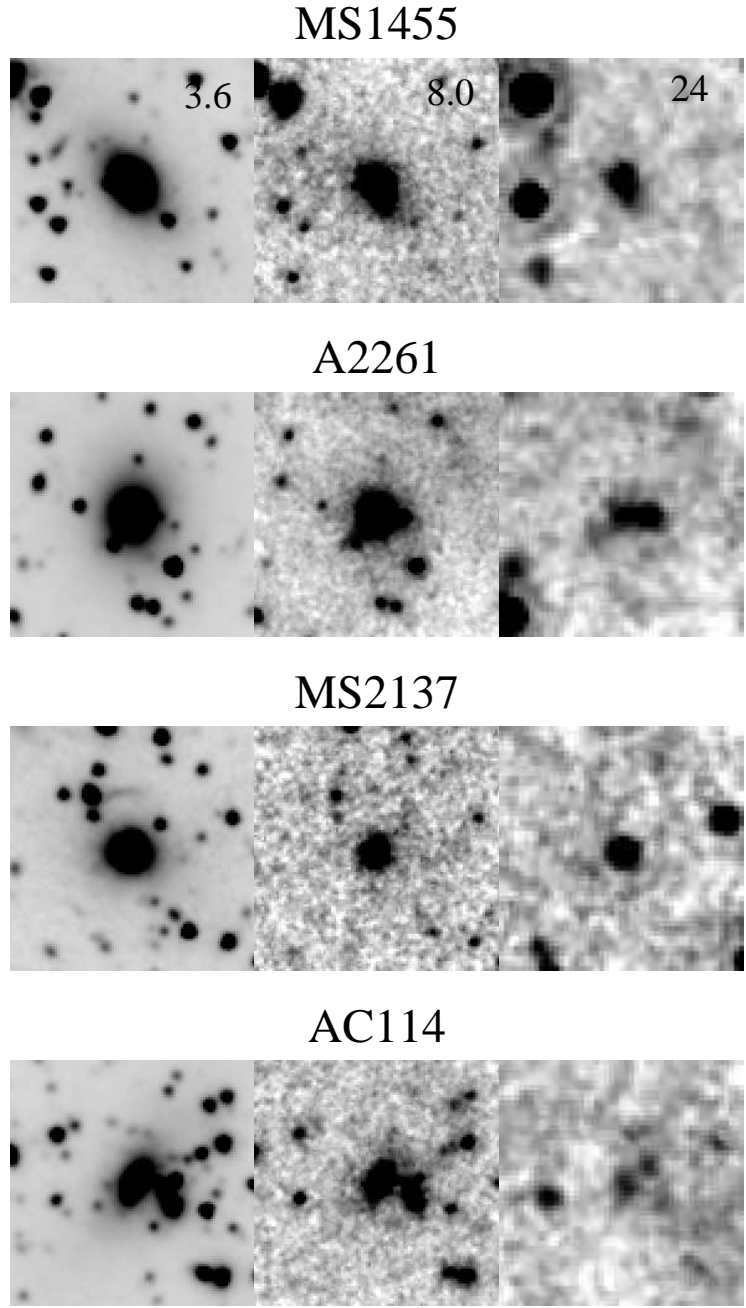}

  \caption{{\em Spitzer} images of the cluster cores for the rest of
  the sample.  From left to right are the 3.6, 8, and 24 $\mu$m
  images, and the clusters are ordered vertically in decreasing 24
  $\mu$m flux densities.  North is up and east is left.  Each image is
  1\arcmin\ on a side, and the pixel scales are same as those in
  Figure~\ref{three_brightest}.
  \label{24um_image}}
\end{figure}

\clearpage

\begin{figure}
  \figurenum{2}

  \epsscale{0.8}

  \hspace*{-2cm}\plotone{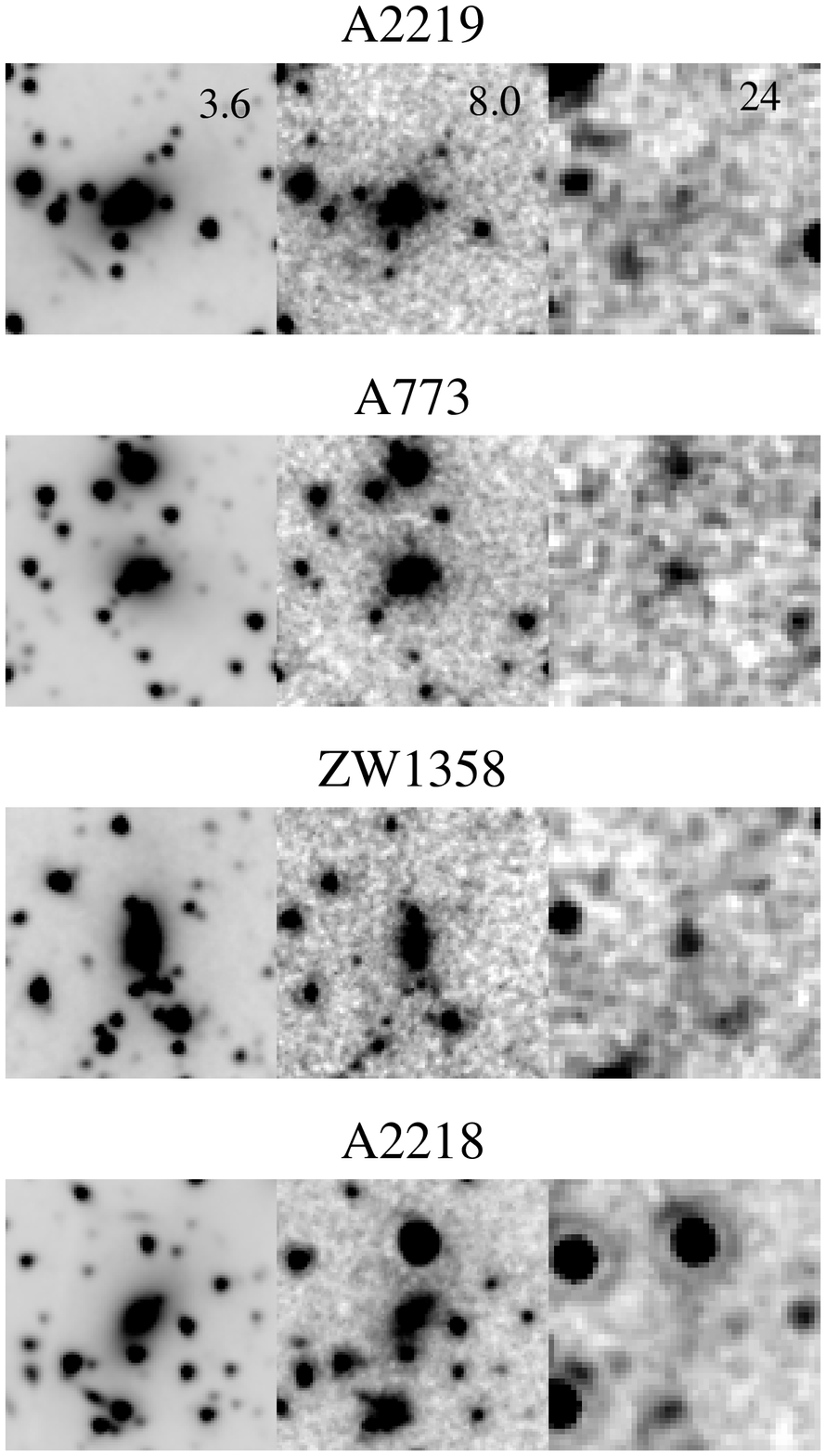}

  \caption{continued.}
\end{figure}

\clearpage

\clearpage

\clearpage

\clearpage

\clearpage

\clearpage

\end{document}